\documentclass[12pt]{article}
\usepackage{a4}
\usepackage{latexsym} 
\usepackage{amssymb}  
\usepackage{amsmath}
\usepackage{graphicx}
\usepackage{epsfig,wrapfig}
\usepackage[ usenames]{color}
\usepackage[rflt]{floatflt}
\begin{document}

\newcommand{\talk}[3]
{\noindent{#1}\\ \mbox{}\ \ \ {\it #2} \dotfill {\pageref{#3}}\\[1.8mm]}
\newcommand{\stalk}[3]
{{#1} & {\it #2} & {\pageref{#3}}\\}
\newcommand{\snotalk}[3]
{{#1} & {\it #2} & {{#3}n.r.}\\}
\newcommand{\notalk}[3]
{\noindent{#1}\\ \mbox{}\ \ \ {\it #2} \hfill {{#3}n.r.}\\[1.8mm]}
\newcounter{zyxabstract}     
\newcounter{zyxrefers}        

\newcommand{\newabstract}
{\clearpage\stepcounter{zyxabstract}\setcounter{equation}{0}
\setcounter{footnote}{0}\setcounter{figure}{0}\setcounter{table}{0}}

\newcommand{\talkin}[2]{\newabstract\label{#2}\input{#1}}

\newcommand{\rlabel}[1]{\label{zyx\arabic{zyxabstract}#1}}
\newcommand{\rref}[1]{\ref{zyx\arabic{zyxabstract}#1}}

\renewenvironment{thebibliography}[1] 
{\section*{References}\setcounter{zyxrefers}{0}
\begin{list}{ [\arabic{zyxrefers}]}{\usecounter{zyxrefers}}}
{\end{list}}
\newenvironment{thebibliographynotitle}[1] 
{\setcounter{zyxrefers}{0}
\begin{list}{ [\arabic{zyxrefers}]}
{\usecounter{zyxrefers}\setlength{\itemsep}{-2mm}}}
{\end{list}}

\renewcommand{\bibitem}[1]{\item\rlabel{y#1}}
\renewcommand{\cite}[1]{[\rref{y#1}]}      
\newcommand{\citetwo}[2]{[\rref{y#1},\rref{y#2}]}
\newcommand{\citethree}[3]{[\rref{y#1},\rref{y#2},\rref{y#3}]}
\newcommand{\citefour}[4]{[\rref{y#1},\rref{y#2},\rref{y#3},\rref{y#4}]}
\newcommand{\citefive}[5]{[\rref{y#1},\rref{y#2},\rref{y#3},\rref{y#4},\rref{y#5}]}
\newcommand{\citesix}[6]{[\rref{y#1},\rref{y#2},\rref{y#3},\rref{y#4},\rref{y#5},\rref{y#6}]}

\begin{center}
{\large\bf Tau Decay Determination of the QCD Coupling}\\[0.5cm]
Antonio Pich\\[0.3cm]
IFIC, Univ. Val\`encia--CSIC, Val\`encia, Spain
\ \& \ Physics Dep., TUM, Munich, Germany
\end{center}

The inclusive character of the total $\tau$ hadronic width renders
possible \cite{BNP:92} an accurate calculation of the ratio
$R_\tau \equiv \Gamma [\tau^- \to \nu_\tau \,\mathrm{hadrons}] /
\Gamma [\tau^- \to \nu_\tau e^- {\bar \nu}_e]$.
Its Cabibbo-allowed component can be written as \cite{PI:2011}
\begin{equation}\label{eq:Rv+a}
 R_{\tau,V+A} \, =\, N_C\, |V_{ud}|^2\, S_{\mathrm{EW}} \left\{ 1 +
 \delta_{\mathrm{P}} + \delta_{\mathrm{NP}} \right\}  ,
\end{equation}
where $N_C=3$ is the number of quark colours and
$S_{\mathrm{EW}}=1.0201\pm 0.0003$ contains the
electroweak radiative corrections. 
The non-perturbative contributions are suppressed
by six powers of the $\tau$ mass \cite{BNP:92}
and can be extracted from the invariant-mass distribution of the final
hadrons \cite{LDP:92a}. From the
ALEPH data, one obtains $\delta_{\mathrm{NP}} \, =\, -0.0059\pm 0.0014$ \cite{DHZ:05}.

The dominant correction ($\sim 20\%$) is the perturbative QCD
contribution \cite{BNP:92} \cite{LDP:92a}
\begin{equation}\label{eq:r_k_exp}
\delta_{\mathrm{P}} \; =\; \sum_{n=1}  K_n \, A^{(n)}(\alpha_s)
\; =\; \sum_{n=1}\,  (K_n + g_n) \, a_\tau^n \;\equiv\;
\sum_{n=1}\,  r_n \, a_\tau^n \, ,
\end{equation}
which is determined by the coefficients of the perturbative expansion
of the ($N_F=3$) QCD Adler function,
already known to $O(\alpha_s^4)$ \cite{BChK:08}:
$K_0 = K_1 = 1$; $K_2 = 1.63982$;
$K_3(\overline{MS}) = 6.37101$ and $K_4(\overline{MS}) =49.07570$.
The functions \cite{LDP:92a}
\begin{equation}\label{eq:a_xi}
A^{(n)}(\alpha_s) \; = \; {1\over 2 \pi i}\,
\oint_{|s| = m_\tau^2} {ds \over s} \,
  \left({\alpha_s(-s)\over\pi}\right)^n\;
 \left( 1 - 2 {s \over m_\tau^2} + 2 {s^3 \over m_\tau^6}
         - {s^4 \over  m_\tau^8} \right)
         \; =\; a_\tau^n + {\cal O}(a_\tau^{n+1})
\end{equation}
are contour integrals in the complex plane, which only depend on
$a_\tau\equiv\alpha_s(m_\tau^2)/\pi$. Using the exact solution
(up to unknown $\beta_{n>4}$ contributions) for $\alpha_s(-s)$
given by the renormalization-group $\beta$-function equation,
they can be numerically computed with very high accuracy~\cite{LDP:92a}.

If the integrals $A^{(n)}(\alpha_s)$ are expanded in powers of $a_\tau$,
one recovers the naive perturbative expansion
of $\delta_{\mathrm{P}}$ shown in the rhs of Eq.~(\ref{eq:r_k_exp}).
This approximation is known as {\it fixed-order perturbation theory} (FOPT), while
the improved expression, keeping the non-expanded values of $A^{(n)}(\alpha_s)$,
is usually called {\it contour-improved perturbation theory} (CIPT) \cite{LDP:92a}.
Even at ${\cal O}(a_\tau^4)$, FOPT gives a rather bad approximation to the
integrals $A^{(n)}(\alpha_s)$, overestimating $\delta_{\mathrm{P}}$ by 12\% at $a_\tau = 0.11$.
The long running of $\alpha_s(-s)$ along the circle $|s|=m_\tau^2$ generates very large $g_n$ coefficients,
which depend on $K_{m<n}$ and $\beta_{m<n}$ \cite{LDP:92a}:
$g_1=0$, $g_2 =  3.56$, $g_3 = 19.99$, $g_4 = 78.00$, $g_5 = 307.78$. These corrections are much larger than the original $K_n$ contributions,
giving rise to a badly behaved perturbative series
(at the four-loop level the expansion of $\alpha_s(-s)$ in powers of $a_\tau$
is only convergent for $a_\tau < 0.11$, which is very close to the physical value of $a_\tau$).
Thus,
it seems compulsory to resum the large logarithms, $\log^n{(-s/m_\tau^2)}$, using the renormalization group. This is precisely what CIPT does.

It has been argued that in the asymptotic regime (large $n$) the renormalonic behaviour of the $K_n$ coefficients could induce
cancelations with the running $g_n$ corrections, which would be missed by
CIPT. In that case, FOPT could approach faster the `true' result provided by the Borel summation
of the full renormalon series. 
This happens actually in the large--$\beta_1$ limit, 
which however does not approximate well the known perturbative series (for $n\le 4$ the true $K_n$ coefficients add constructively with the $g_n$ contributions).
%
%
Models of higher-order corrections which assume a precocious asymptotic behaviour of the Adler function already at $n=3,4$ \cite{BJ:08} \cite{CF:10} seem to favour the FOPT result.
The CIPT procedure is much more reliable in all other scenarios.

The present experimental value $R_{\tau,V+A} = 3.4771\pm 0.0084$ \cite{HFAG} implies
$\delta_{\mathrm{P}} = 0.2030 \pm 0.0033$.
The two different treatments of the perturbative series result in
\begin{eqnarray}\label{eq:alpha-result}
\alpha_s(m_\tau^2)_{\mathrm{CIPT}} & =& 0.3412 \pm 0.0041_{_{\delta_{\mathrm{P}}}} \, {{}^{+\, 0.0069}_{-\, 0.0064}}_{{K_5}}
\, {{}^{+\, 0.0050}_{-\, 0.0001}}_{_\mu}
\, {{}^{+\, 0.0039}_{-\, 0.0034}}_{_{\beta_5}}
\; =\; 0.344 \pm 0.014\, ,
\\
\alpha_s(m_\tau^2)_{\mathrm{FOPT}} & = &
0.3194 \pm 0.0028_{_{\delta_{\mathrm{P}}}} \, {{}^{+\, 0.0039}_{-\, 0.0035}}_{{K_5}}
\, {{}^{+\, 0.0105}_{-\, 0.0045}}_{_\mu}\, {{}^{+\, 0.0019}_{-\, 0.0045}}_{_{\beta_5}}
\; =\; 0.321 \pm 0.015\, .
\end{eqnarray}
Higher-order corrections have been estimated
adding the fifth-order term $ K_5\, A^{(5)}(\alpha_s)$ with $K_5 = 275\pm 400$.
We have also included the 5-loop variation with changes of the renormalization scale in the range $\mu^2/(-s) \in [0.4,2.0]$.
The error induced by the truncation of the $\beta$ function at fourth order has been conservatively estimated
through the variation of the results at five loops, assuming $\beta_5 =\pm \beta_4^2/\beta_3 = \mp 443$;
in CIPT this slightly changes the values of $A^{(n)}(\alpha_s)$, while in FOPT it increases the scale sensitivity.
The FOPT result shows as expected \cite{LDP:92a} \cite{ME:09} a much more sizeable $\mu$ dependence, but it gets smaller errors from $\delta_{\mathrm{P}}$ and $K_5$.
The three theoretical uncertainties
($K_5$, $\mu$, $\beta_5$) have been added linearly and their sum  combined in quadrature with the
`experimental' error  from $\delta_P$.

Combining the two results with the PDG prescription (scale factor $S=1.14$), one gets
$\alpha_s(m_\tau^2)= 0.334 \pm 0.011$. We keep conservatively the smallest error, i.e.
\begin{equation}
\alpha_s(m_\tau^2)\; =\; 0.334 \pm 0.014
\qquad\quad\longrightarrow\qquad\quad
\alpha_s(M_Z^2)\; =\; 0.1204 \pm 0.0016
\, .
\end{equation}
The resulting value is in excellent agreement with the direct measurement of $\alpha_s(M_Z)$ at the $Z$ peak, providing a very significant experimental verification of
{\it asymptotic freedom}.


%

\end{document}